\documentclass[graybox]{svmult}
\usepackage{type1cm,graphicx,natbib,amsmath, amssymb, amsfonts,charter,enumitem,color,xcolor,booktabs,caption,placeins} 

\def\bone{\mathbf{1}}\def\bzero{\mathbf{0}} 
\def\D{\mathbf{D}} \def\E{\mathbf{E}} \def\I{\mathbf{I}} \def\V{\mathbf{V}}
\def\f{\mathbf{f}}  \def\m{\mathbf{m}}
\def\s{\mathbf{s}} \def\x{\mathbf{x}}\def\y{\mathbf{y}}
\def\bepsilon{{\boldsymbol{\epsilon}}}\def\btau{{\boldsymbol{\tau}}}
\def\cD{{\mathcal D}}\def\cM{{\mathcal M}}
\newcommand{\RR}{\mathbb{R}}
\def\seq#1#2{#1{:}#2}
\def\eqn#1{eqn.~(\ref{eq:#1})} 
\def\diag{\textrm{diag}}
\def\beq#1{\begin{equation}\label{eq:#1}}\def\eeq{\end{equation}}

\def\portar{r^*}
\def\mc#1{\multicolumn{1}{c}{#1}}
\def\msp{\qquad\phantom{.}\qquad}

\begin{document}

\title*{Predictive Decision Synthesis for Portfolios: Betting on Better Models}
\titlerunning{Portfolio BPDS: Betting on Better Models} 
\author{{\large Emily Tallman and Mike West}\\ \phantom{.} \\  
             Department of Statistical Science\\  Duke University \\ Durham NC 27708. U.S.A.\\ \phantom{.} \\ 
             {\normalsize\email{emily.tallman@duke.edu, mike.west@duke.edu}}}
\authorrunning{Tallman \& West}
\maketitle
{\small\today\bigskip\bigskip}

\abstract{We discuss and develop Bayesian dynamic modelling and predictive decision synthesis for portfolio analysis. The context involves model uncertainty with a set of candidate models for financial time series with main foci in sequential learning, forecasting, and recursive decisions for portfolio reinvestments. The foundational perspective of Bayesian predictive decision synthesis (BPDS) defines novel, operational analysis and resulting predictive and decision outcomes. A detailed case study of BPDS in financial forecasting of international exchange rate time series and portfolio rebalancing, with resulting BPDS-based decision outcomes compared to traditional Bayesian analysis, exemplifies and highlights the practical advances achievable under the expanded, subjective Bayesian approach that BPDS defines.} 

 
\setcounter{page}{0}\thispagestyle{empty}\newpage

\section{Introduction}
\label{sec:Intro}

 Portfolio optimization is a major area of application and success of Bayesian decision analysis, going back at least to~\cite{Markowitz} and arguably earlier~\citep{deFinetti1940,Markowitz2006,Rubinstein2006}.  More recent Bayesian literature has emphasized financial time series modelling and forecasting to link with portfolio decision analysis~(e.g.~\citealp{Quintana1987};~\citealp{WestHarrisonYellowBook1stEdition}, chap. 15;~\citealp{WestHarrisonYellowBook19972ndEdition}, chap. 16;~\citealp{Aguilar2000};~\citealp{KARLSSON2013791};~\citealp{ZhaoXieWest2016ASMBI}; Gruber and West, 2016 and 2017;\nocite{GruberWest2016BA,GruberWest2017ECOSTA}~\citealp{PradoFerreiraWest2021}, chap. 10), along with continuing concern for customizing utility specifications to guide Bayesian portfolio selection~\citep[e.g.][and references therein]{IrieWest2018portfoliosBA}

Here we explicitly address portfolio decision goals when multiple forecasting models are to be explored and combined.  Accounting for model uncertainty in pure forecasting is routine~\citep[e.g.][]{WestHarrisonYellowBook19972ndEdition,Karlsson2004,andersson_bayesian_2008,Cheng2012, Gold2015, WANG2016136, Steel2020, LavineLindonWest2021avs, Forecastcombination, BernaciakGriffin2024}, but decision goals are not generally incorporated in these statistical approaches. In portfolio analysis, the main goal of forecasting asset returns is selecting desirable portfolios; we argue that this should be a key focus in combining relevant forecasting models. Recent Bayesian “goal-focused” methods aim to incorporate specific prediction goals beyond those of traditional Bayesian model averaging~\citep[e.g.][]{Karlsson2007, Pettenuzzo2016, LavineLindonWest2021avs, LoaizaMayaJOE2021, West2020Akaike,  McAlinn2021, BernaciakGriffin2024}. However, this literature rarely addresses explicit decision goals. A related literature concerns combining portfolio rules~\citep[e.g.][]{Kan2007, Demiguel2009, TU2011204, portfoliorules}, though without explicit consideration of forecasting models. 

Bayesian Predictive Decision Synthesis (BPDS), introduced in~\cite{TallmanWest2023}, closes these gaps. BPDS is a fully Bayesian, theoretically founded framework to evaluate, compare, and combine sets of candidate models based on both their anticipated and historical decision outcomes as well as pure predictive performance and validity.  We develop BPDS here in the setting of financial forecasting for portfolio decisions.
BPDS is an expansion of Bayesian Predictive Synthesis (BPS)~\citep{McAlinnWest2018, McAlinnEtAl2019, McAlinn2021,JohnsonWest2022}, with explicit incorporation of the decision context and goals.  BPDS aims to more highly weight models and outcomes that are better-performing in both recommended (Bayesian decision-theoretic) optimal decisions and more traditional statistical forecasting accuracy metrics. 
The context of sequential portfolio decision analysis is a canonical setting for the development and exploitation of BPDS for improved predictive decision-making in the central practical context of model uncertainty. 

\section{Bayesian  Predictive Decision Synthesis\label{sec:score_funct}} 

\subsection{Model Uncertainty and Portfolio Decisions}

Quantities of interest and relevant notation are defined as follows, indexed by time $t$ and implicitly representing the process of sequential analysis as time progresses.
This is a standard setting of Bayesian model uncertainty analysis, indexed by time. 
We use the standard notation that $\cD_{t-1}$ represents all information and data accrued up to and including time $t-1$, the point at which a decision is to be made to target some outcome at time $t$. Relevant quantities include:
\begin{itemize}
\item A forecast outcome $q-$vector $\y_t$. Examples are the time $t$ daily returns on $q$ assets, representing the percent change in price from time $t-1$ to time $t$, or a set of returns over a multi-period path ahead from times $t$ to $t+h$ for some portfolio horizon $h.$ 
\item  An explicit, decision problem targeting the outcome at time $t$, in this case selecting a set of portfolio weightings $\x_t$ on the $q$ assets. 
\item  A set of $J$ models $\cM_j$, $j=\seq1J$, such as a set of dynamic regression models for the asset return vector with different predictors across models, or other distinct model forms. 
\item Initial model probabilities $\pi_{tj} = Pr(\cM_j|\cD_{t-1})$, such as from traditional Bayesian model averaging (BMA) based on prior data and information, and/or from historical performance in goal-focused forecasting such as Bayesian adaptive variable selection (AVS:~ \citealp{LavineLindonWest2021avs}, a special case of BPS)-- or related methods~\citep[e.g.][]{LoaizaMayaJOE2021}.
\item  Predictive densities provided by each model $\cM_j$, $p_{tj}(\y_t|\cM_j, \cD_{t-1})$. For example, a one-step predictive for that day's FX returns, or a multi-step predictive for daily returns over the next $h$ days.
\item Model $\cM_j$ applies decision analysis with its utility function $u_{tj}(\y_t,\x_t)$ to evaluate the model-specific, optimal portfolio $\x_{tj}$ for time $t$. This would be the optimal portfolio decision if the decision-maker restricted the analysis only to include $\cM_j$ and its associated utility function. 
\end{itemize}
Faced with this model uncertainty question, the decision-maker aims to synthesize information across models to forecast $\y_t$ and decide on a final portfolio weight vector $\x_t$ under whatever overall utility function is chosen. This is then repeated sequentially over time. 

\subsection{BPDS}
In the above setting, BPDS has the following ingredients.  This follows and extends the main methodological developments from the BPDS theory in~\cite{TallmanWest2023}.
 
\subsubsection{Initial Mixture and Baseline Model} 
BPDS requires a \lq\lq baseline'' predictive density  $p_0(\y_t|\cM_0, \cD_{t-1})$ based on a notional additional model $\cM_0$ and a corresponding baseline optimal decision $\x_{t0}.$  The baseline serves as an over-dispersed \lq\lq safe haven" model whose predictions more heavily favor regions of the $\y_t$ outcome space than are less well-supported by the initial mixture of the $J$ models.  The initial probabilities $\pi_{tj}$ are extended and renormalized to include a non-zero $\pi_{t0}$-- an initially \lq\lq small" probability on $\cM_0,$ representing the concern that \lq\lq all (initial $J$) models are wrong", i.e., the issue of model set incompleteness. 

We thus have the density 
$p_t(\y_t,\cM_j|\cD_{t-1}) = \pi_{tj} p_{tj}(\y_t|\cM_j, \cD_{t-1})$ defined over models $j=\seq 0J$ and outcomes $\y_t$ jointly. The margin for $\y_t$, simply 
$p_t(\y_t|\cD_{t-1}) = \sum_{j=\seq 0J}\pi_{tj} p_{tj}(\y_t|\cM_j, \cD_{t-1}),$ 
is the called the {\em initial mixture}.    

This framework reduces to the traditional (sequential, dynamic) Bayesian model uncertainty analysis when $\pi_{t0}=0$ and if the other $\pi_{tj}$ are 
appropriately based on historical data~\citep[][Sect.~12.2]{WestHarrisonYellowBook19972ndEdition}. In that special case, the initial mixture 
$p_t(\y_t|\cD_{t-1}) $ is simply that based on standard BMA.

\subsubsection{BPDS Model Probabilities and Densities}  
BPDS modifies the initial density $p_t(\y_t,\cM_j|\cD_{t-1})$ to an updated BPDS density 
$f_t(\y_t,\cM_j|\cD_{t-1}) \propto  \alpha_{tj}(\y_t) p_t(\y_t,\cM_j |\cD_{t-1})$
where the $\alpha_{tj}(\y_t)$ are positive, model and outcome-dependent weights. That is, 
\beq{BPDSmix}
f(\y_t,\cM_j|\cD_{t-1})  
=      k_t\pi_{tj}\alpha_{tj}(\y_t) p_{tj}(\y_t|\cM_j, \cD_{t-1}), \quad \y_t\in \RR^q, \ j=\seq0J,
\eeq
with $k_t$ normalizing over both outcomes and models, i.e., 
\beq{BPDSajandk} 
k_t^{-1} = \sum_{j=\seq0J} \pi_{tj} a_{tj}
\quad\textrm{where}\quad 
a_{tj} = \int_{\y_t\in\RR^q} \alpha_{tj}(\y_t) p_j(\y_t|\cM_j, \cD_{t-1})d\y_t.
\eeq
The implied margin for $\y_t$ is the {\em BPDS mixture}  
\begin{equation}
\label{BPDSmixpredy}
    f(\y_t|\cD_{t-1})  =
   \sum_{j=\seq0J} \tilde\pi_{tj} f_{tj}(\y_t|\cM_j, \cD_{t-1})
\end{equation}
where, for each  model $j=\seq0J,$  
\begin{equation}\label{BPDSfjpij} 
     f_{tj}(\y_t|\cM_j, \cD_{t-1}) = a_{tj}^{-1} \alpha_{tj}(\y_t) p_j(\y_t|\cM_j, \cD_{t-1})\quad\textrm{and}\quad
\tilde\pi_{tj} = k_t \pi_{tj} a_{tj}. 
\end{equation}
The $f_{tj}(\cdot|\cdot)$ are {\em  BPDS model predictive densities}, 
the $\tilde\pi_{tj}$ are {\em BPDS model probabilities}, and the mixture 
$f(\y_t|\cD_{t-1})$ is the {\em BPDS predictive density}.

\subsubsection{Portfolio BPDS Scores and Relaxed Entropic Tilting}
BPDS model weights are generally and uniquely defined as 
\beq{alphajET}\alpha_{tj}(\y_t) = \exp\{ \btau_t' \s_{tj}(\y_t,\x_{tj})\}\eeq with ingredients discussed below. 
The model-specific optimal portfolio vectors $\x_{tj}$ play critical roles in these BPDS model weightings.  
 
First, $\s_{tj}(\y_t,\x_{tj})$ is a vector of {\em scores} depending on outcome $\y_t$ and the optimal portfolio $\x_{tj}$  under $\cM_j$. Here $\x_{tj}$ is generated by optimizing whatever utility function is chosen for the decision analysis under $\cM_j$.  
The elements of $\s_{tj}(\y_t,\x_{tj})$ are realized values of $k$ utility functions at the $\cM_j-$optimal portfolio $\x_{tj}$ were  $\y_t$ to be the actual outcome. These utility functions are chosen so that higher scores are desirable. This allows for scoring models based on multiple metrics, expanding to multi-criteria decision analysis. Restrictions on score functions depend on the context and models involved. An example of a class of relevant score functions is given in the following section.

Second, $\btau_t$ is a {\em BPDS tilting vector} that defines relative weights and the directional impact of changes in score elements. Larger values in $\btau_t$ lead to more appreciable BPDS weightings, while $\btau_t = \bzero$ recovers the prior mixture.  The term tilting relates to the Bayesian decision-theoretic foundation and derivation of~\eqn{alphajET} which is based on relaxed entropic tilting (RET: see~\citealp{TallmanWestET2022}; 2023, 
section 3;~\citealp[][sect.~2.3]{West2023constrainedforecasting}). 

The specification of $\btau_t$ references expected scores. 
Taking expectations over $(\y_t,\cM_j)$ jointly, 
the initial model $p_t(\y_t,\cM_j|\cD_{t-1})$ has {\em initial expected score}  
$E_p[\s_{tj}(\y_t,\x_{tj})|\cD_{t-1}].$   High scores are desirable. The RET framework aims to achieve scores higher than initial, asking for the BPDS distribution 
 $f_t(\y_t,\cM_j|\cD_{t-1})$ to have expected score {\em at least}
 $E_p[\s_{tj}(\y_t,\x_{tj})|\cD_{t-1}] + 
 \bepsilon_t$ for some non-negative $k-$vector $\bepsilon_t$ with at least one positive element. With this constraint, RET chooses $f(\cdot|\cdot)$ to minimize the 
 the K\"ullback-Leibler (KL) divergence of $p(\cdot|\cdot)$ from $f(\cdot|\cdot)$, yielding two key results: (i) $f(\cdot|\cdot)$ is defined in~\eqn{BPDSmix} with weighting function precisely as in~\eqn{alphajET}; (ii) $f(\cdot|\cdot)$ achieves {\em exactly} the
 expected score improvement bound specified by $\bepsilon_t$, i.e., 
 $E_f[\s_{tj}(\y_t,\x_{tj})|\cD_{t-1}] = E_p[\s_{tj}(\y_t,\x_{tj})|\cD_{t-1}] + 
 \bepsilon_t.$  
The RET theory shows that any feasible choice of  $\bepsilon_t$ corresponds to a unique $\btau_t$. The decision-maker has the liberty to choose $\bepsilon_t$ in each specific portfolio context; various 
choices are explored in the portfolio examples below. 

\section{Risk{\,}:{\,}Return Score Functions for Portfolios}\label{sec:score_funct_sub}
The section temporarily drops the time index $t$ from the notation, for clarity. Thus $\y$ is a vector of asset returns, and $\s_j(\y,\x_j)$ a vector score on $\cM_j$ using model-specific portfolio vector $\x_j$ were $\y$ to be the observed outcome.

\subsection{Utility Theory, Bivariate Scores and Risk Tolerance} 
Consider $k=2$ score dimensions and the class of vector scores 
$\s_j(\y, \x_j) =  (e_j, -e_j^2/2)' $ where $e_j=r_j-\portar$ with {\em portfolio return} 
$r_j = \x_j'\y$ and a chosen {\em target portfolio return} $\portar$. 

These bivariate scores are motivated by 
the classic, bounded-above risk-averse exponential utility function for return $r$ given by 
$U(r) = -\exp (-r/d)$ for some risk tolerance level parameter $d>0$. A second-order Taylor series approximation around $r=\portar$ gives $U(r)\approx U_q(r)$ with quadratic function
$U_q(r) = U(\portar)\{1-e/d+e^2/(2d^2)\}$ and $e=r-\portar.$ This is a very accurate approximation across ranges of $r,\portar$ that correspond to realistic models for return prediction in portfolio analyses, especially for daily returns.  Up to constants, $U_q(r)$ essentially reduces to the Markowitz function 
 $e-e^2/(2d)$,  maximized at $r=\portar+d$. Were we to consider BPDS based on only a univariate score, then $s_j(\y,\x_j) = e_j - e_j^2/(2d)$ is faithful to this natural utility function and theoretical considerations. This has the contextual interpretation of $d$ as a chosen level of risk tolerance. The implied BPDS weight function would then be  
 $\alpha_j(\y, \x_j) = \exp\{\tau (e_j -e_j^2/(2d))\}$.  
 
In contrast, choosing the suggested and more general bivariate score $\s_j(\y, \x_j) =  (e_j, -e_j^2/2)' $  allows flexibility in addressing risk tolerance. This leads to 
 $\alpha_j(\y, \x_j) = \exp\{\tau_1 e_j -\tau_2 e_j^2/2\}$ which agrees with the univariate case above if/when  $\tau = \tau_1$ and $d= \tau_1/\tau_2$; this defines a maximizing target return $\portar + d$ at this level of risk tolerance. Hence the bivariate score allows more flexibility in the role of the return/risk elements and their impact on the BPDS analysis while overlaying and replicating the classical exponential utility-based example as a special case.  Operationally, it also yields the direct interpretation of $\tau_1/\tau_2$ as the implied level of risk tolerance. 

\subsection{Bivariate Scores and Tilting Vectors}
In this section, we step aside from the mixture context to focus on generating portfolio-relevant insights from RET applied to a single model. This aids in understanding the roles of implied tilting vectors in the portfolio context and the relation to risk tolerance, among other details. 

For this, consider predictions from any initial model $\y \sim p(\y)$ such that the implied return $r=\x'\y \sim p(r)$ has mean $E_p(r)=\portar+f$, where $f$ can be positive or negative, and variance $V_p(r)=q.$  We highlight aspects of the implied distribution of the bivariate score $\s(\y,\x) =  (e, -e^2/2)'$ involving excess return $e=r-\portar,$ and practical implications for BPDS analysis.

\subsubsection{Distributions of Scores} \label{sec:scoredistributions}
Whatever the distribution of $r$ and the score vector under $p(\y)$ may be, knowing only 
$E_p(r)=\portar+f$ and $V_p(r)=q$  implies $E_p[\s(\y,\x)] = (f, -(f^2+q)/2)'.$ 
To explore the distribution of $\s(\y,\x)$ further, useful insights arise in 
assuming a normal approximation to $p(r)$. 

With daily returns on FX and stock indices, $f$ and $q^{1/2}$ are typically of the order of 0.01 or less; this is relevant in exploring score uncertainty via the variance matrix $\V_p= V_p[\s(\y,\x)]$.
Were $p(r)$ to be normal, then it is easy to show that 
$V_p(-e^2/2) = (f^2+q/2) V_p(e).$ Hence, $e$ is more uncertain than $-e^2/2$ 
if $f^2+q/2<1$. With practical levels of small \% returns,  this is very likely, and $V_p(-e^2/2)$ is typically much smaller than $V_p(e)$. That is, we expect to be much more uncertain about the first score element than the second.

While the two score elements are deterministically related, their joint distribution is not wholly degenerate since knowing $-e^2/2$ does not precisely determine $e$. 
In fact, the covariance is easily be shown to be  
$C(e,-e^2/2) = -fq$. The implied score correlation is then
$\rho \equiv Cor(e,-e^2/2) = -f_0(f_0^2+0.5)^{-1/2}$ where $f_0=f/\sqrt{q}.$
Note that: (i) $\rho$ has the sign of $-f$ and depends only on $f_0,$ the expected excess return in standardized $N(0,1)$ units; (ii) $\rho$ is a decreasing function of $f_0$ and $|\rho|$ is an increasing function of $f_0$; (iii) over the range $f_0=-1$ to $f_0=1$,  the correlation decreases from roughly 0.8 to $-0.8$; (iv) for any given $q,$ the absolute value $|\rho|$ is an increasing function of $|f|$; (v) for any given $f$, the value $|\rho|$ is a decreasing function of $q$.

\subsubsection{RET Distribution and Tilting Vector in the Normal Example}
Suppose that $p(\y)$ is normal.  For a given tilting vector $\btau=(\tau_1,\tau_2)'$ and any $\x,$ the resulting RET distribution $f(\y) \propto 
 \exp\{\btau'\s(\y,\x)\}  p(\y) $ is then also normal with easily computed moments. The implied $f(r)$ on the return $r=\x'\y$ alone is then normal 
with moments that are easily shown to be  
$$E_f(r) = \portar + (f+q\tau_1)/(1+q\tau_2) \quad\textrm{and}\quad V_f(r) = q/(1+q\tau_2).$$
Hence the expected score $\m= (m_1,m_2)'$ under $f(\cdot)$ has elements
\begin{align}\label{eq:normalportegmeans}
\begin{split} 
     m_1 &= E_f(r-\portar) = (f+q\tau_1)/(1+q\tau_2), \\
     m_2 &= -E_f[(r-\portar)^2]/2 =  -\{q+(1+q\tau_2)m_1^2\}/\{2(1+q\tau_2)\}.
\end{split}
\end{align}
These equations are easily solved for $\btau$ as a function of $\m$ to give {\em analytic} evaluation of the tilting vector. This can be shown to give
$$
\tau_1 = \frac{-m_1q-m_1^2f-2m_2}{q(m_1^2+2m_2f)} \quad\textrm{and}\quad 
\tau_2 = \frac{-m_1^2-2m_2-q}{q(m_1^2+2m_2)}.
$$
With $m_1>f$ and $\tau_1>0$ it follows that $\tau_2>0$ as is desirable. Then \eqn{normalportegmeans} leads to $\tau_1 > \tau_2 f$  or $d>f$ where $d=\tau_1/\tau_2$ has, as noted earlier, interpretation in terms of risk tolerance. We also easily have that  
\begin{equation}\label{eq:normalportegtauonetautwo}d =  \tau_1/\tau_2 = (m_1q+m_1^2f+2m_2f)/(m_1^2+2m_2+q). 
\end{equation}
Some examples below choose $\portar$ so that $E_p(r) = \portar$, i.e., $f = 0$ in the above. In that case,~\eqn{normalportegtauonetautwo} reduces to the simple form
$d=  m_1q/(m_1^2+2m_2+q).$ This special case leads to $\rho = 0$, i.e., the two score dimensions are uncorrelated. 

The above expressions for $d$ relate to our earlier comment about the bivariate score allowing the analysis to more flexibly reflect actual attitudes to risk through the direct specification of expected target scores $(m_1,m_2)',$ any choice
of which directly determines risk tolerance.  Some examples follow. 

\subsubsection{Examples: Tilting Vectors and Risk Tolerance}
Numerical examples highlight the relationships between taget expected scores and the implied tilting vector $\btau$ and associated risk tolerance $d=\tau_1/\tau_2.$
\begin{figure}[b!]
    \centering
    \includegraphics[width=.7\textwidth]{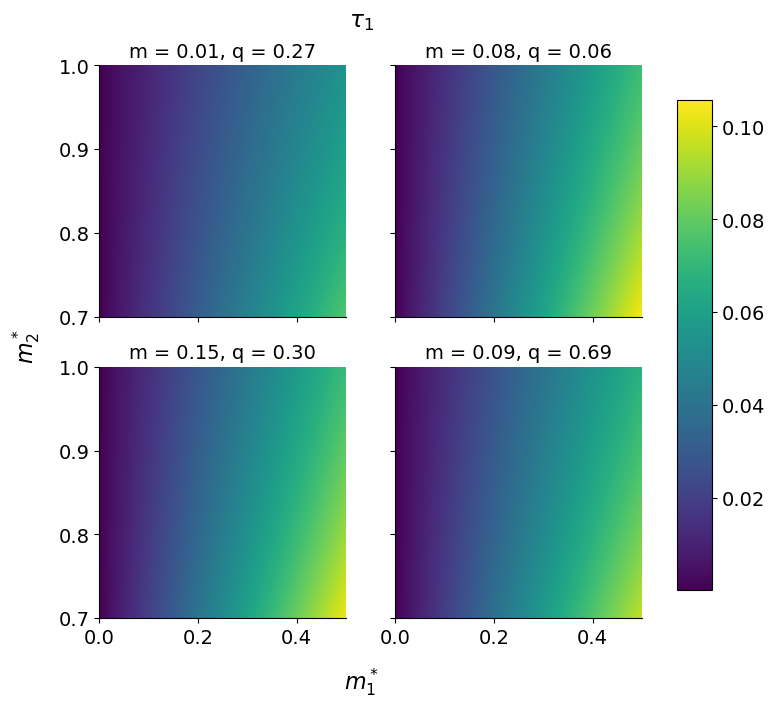}
    \caption[$\tau_1$ in normal portfolio example]{$\tau_1$ across ranges of score improvement percentages $m_1^*$ and $m_2^*$, and $m = E(r), q = V(r)$.}
    \label{fig:tau1}
\end{figure}

\begin{figure}[hbtp!]
    \centering
    \includegraphics[width=.7\textwidth]{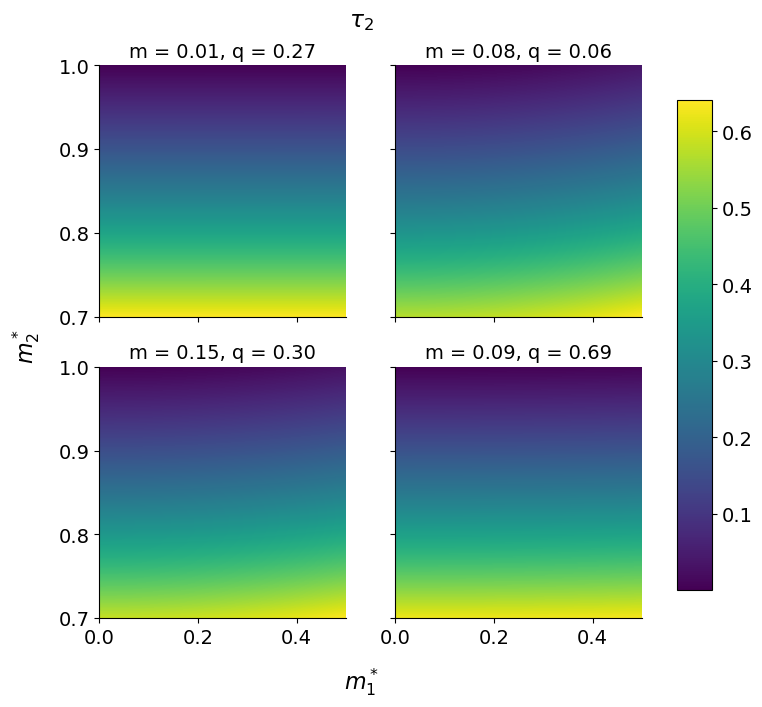}
    \caption[$\tau_2$ in normal portfolio example]{$\tau_2$ across ranges of score improvement percentages $m_1^*$ and $m_2^*$, and $m = E(r), q = V(r)$.}
    \label{fig:tau2}
\end{figure}

\begin{figure}[hbpt!]
    \centering
    \includegraphics[width=.7\textwidth]{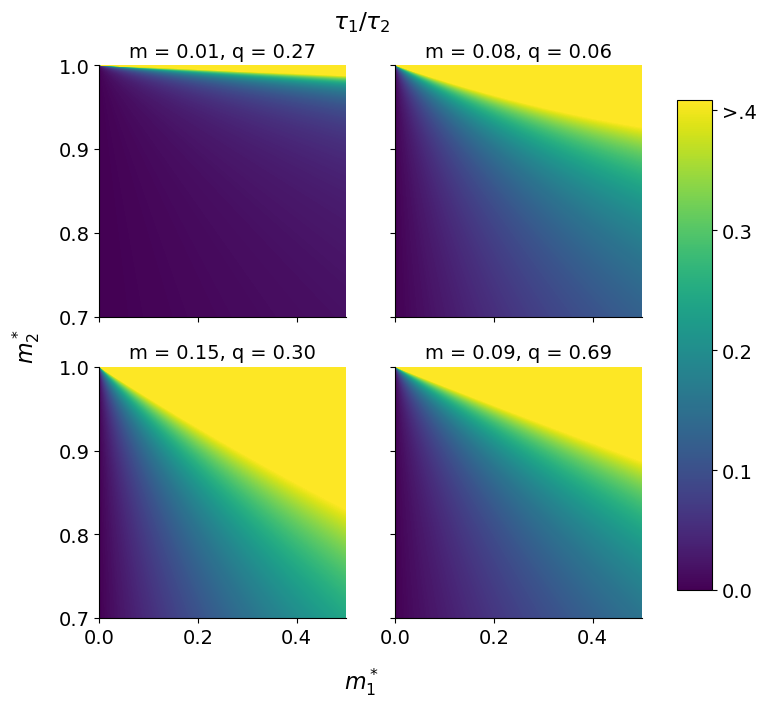}
    \caption[$\tau_1/\tau_2$ in normal portfolio example]{$\tau_1/\tau_2$ across ranges 
      of score improvement percentages $m_1^*$ and $m_2^*$, and $m = E(r), q = V(r)$.}
    \label{fig:tau_ratio}
\end{figure}

Specifying target expected scores $\m$ under $f(\cdot|\cdot)$ can be simply addressed through utilizing percent improvements over the initial expectations under $p(\cdot|\cdot).$ Figures~\ref{fig:tau1},~\ref{fig:tau2} and~\ref{fig:tau_ratio} 
display values of $\tau_1$, $\tau_2$ and $d=\tau_1/\tau_2$ for a few practical values of $m = \portar = E_p(r)$ and $q=V_p(r)$ and against different levels of percent target improvements in each of the score dimensions; the percent levels are denoted by $m_1^*$ and $m_2^*$, where $100m_1^*$ is the percent improvement in expected return over $m$ and $100m_2^*$ is the percent improvement in $-E_p[(r-\portar)^2]/2$.  The general lack of correlation between the two score dimensions is due to the lack of prior score correlation. Then, the risk tolerance level $\tau_1/\tau_2$ may become large for small values of $m_2^*$ and large values of $m_1^*$, indicating that these values should be carefully set to properly constrain risk.

\section{Case Study: Data, Models and Portfolio Construction} 

\subsection{Setting and Data}

The study involves an extension of the example applying BPDS to portfolio analysis in \cite{TallmanWest2023}, with more recent data, adapted models, and an in-depth investigation of the BPDS hyperparameters used.

\begin{figure}[t!]
    \centering
    \includegraphics[width=0.8\textwidth]{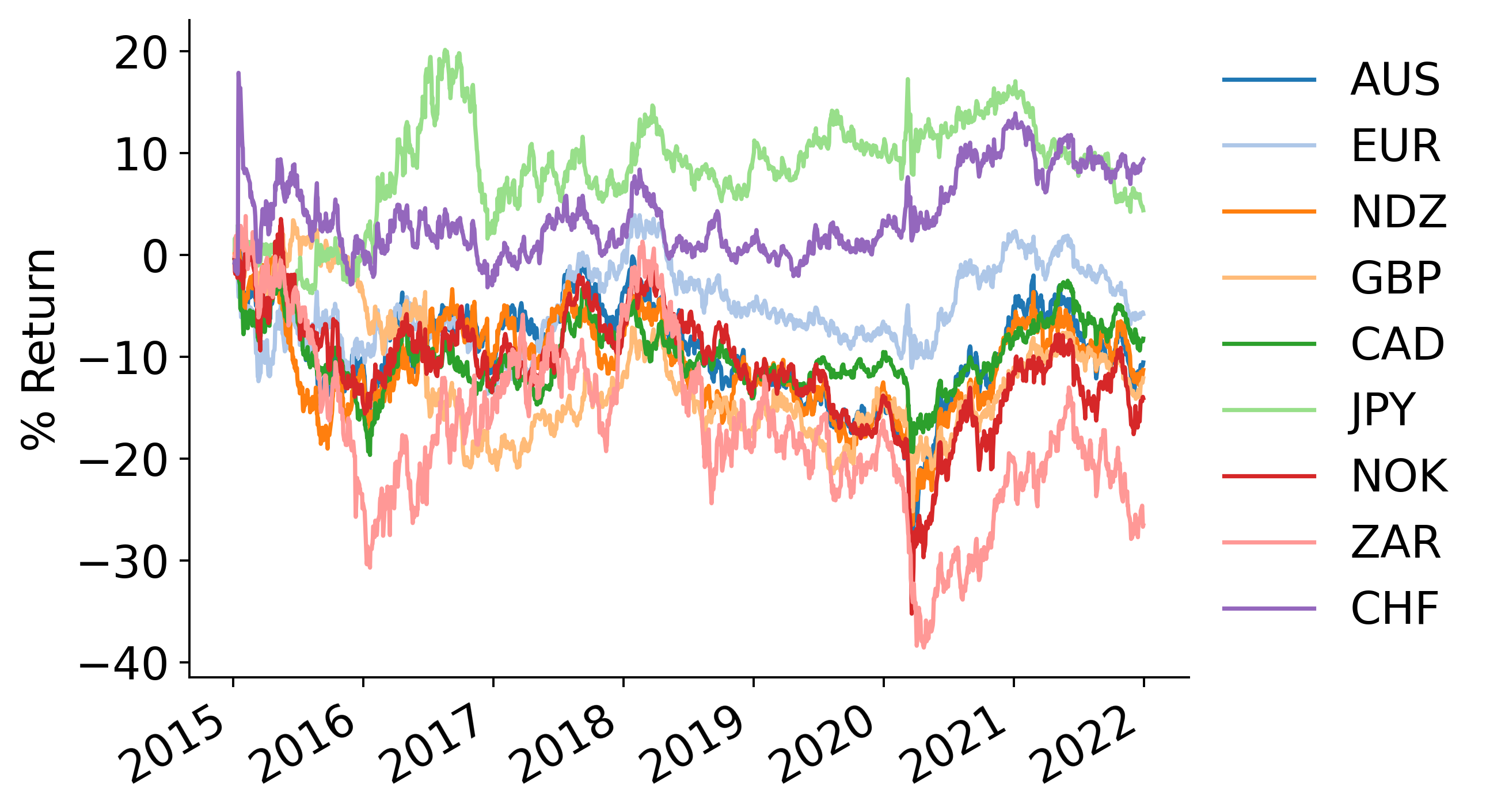}
    \caption{Cumulative FX returns for the $q=9$ currencies.}
    \label{fig:curr_returns}
\end{figure}

The data set includes daily returns for $q=9$ currencies beginning in January 2001 and ending in December 2021. The motivation for this asset set is to ensure model performance is not anchored in the performance of the market indices used in \cite{TallmanWest2023}, with the goal of providing additional value beyond the market return. The $q=9$ FX assets are noted in Table \ref{tab:asset_names2} with cumulative returns over the time period in Figure~\ref{fig:curr_returns}.

\begin{table}[b!]
\centering
\begin{tabular}{@{}llcll@{}}
\em Ticker\qquad & \em Currency & \msp\msp & \em Ticker\qquad & \em Currency \\
\midrule
AUD    & Australian Dollar      && JPY    & Japanese Yen           \\
EUR    & Euro                   && NOK    & Norwegian Krone        \\
NZD    & New Zealand Dollar     && ZAR    & South African Rand     \\
GBP    & UK Pound Sterling      && CHF    & Swiss Franc            \\
CAD    & Canadian Dollar        && & \\
\end{tabular}
\caption{FX series in the portfolio case study.}
\label{tab:asset_names2}
\end{table}

\subsection{Models and Model-Specific Portfolios} 

The $\cM_j$  are time-varying vector autoregressive (TV{--}VAR) models applied to log FX prices. Details of these standard models are in~\citet[][Appendix]{TallmanWest2023}  with full theory in~\citet[][sect.~10.8]{PradoFerreiraWest2021}. On day $t-1$ each model predicts the vector  $\log(\mathbf{p}_t)$ where $\mathbf{p}_t$ is the vector of asset prices on day $t$. These transform to returns via $\y_t = \mathbf{p}_{t}/\mathbf{p}_{t-1} - 1$. 

The dataset is divided into 3 time periods. The period to the beginning of 2015 is used to fit a large set of models, the next 4 years (2015-2018 inclusive) is used to select a subset of \lq\lq better-performing models'', and the final 3 years (2019-2021 inclusive) defines a hold-out data test period.
The larger initial set contains models of AR order $p \in \{1, 2, 3\}$, reflecting potential $1{-}3$ day momentum effects in FX prices. In each model, each univariate series $y_{ti}$ is predicted by all values of $\y_{t-p:t-1}$ with differing coefficients for each asset. These coefficients evolve over time as multivariate random walks; each model uses a state evolution discount factor $\delta = 0.9995$ to govern the degree of these changes. The model set is further expanded with a range of possible discount factors $\beta$ governing the evolution of the time-varying volatility matrix in each model~\citep[][chaps.~9~\&~10]{Aguilar2000,IrieWest2018portfoliosBA,PradoFerreiraWest2021}. Discount factor values $\beta\in \{.94, .98, .995\}$  define differing degrees of change in levels of volatility of each of the assets over time, as well as-- critically for portfolio analysis-- in the inter-dependencies among the assets as represented by time-varying covariances, with smaller values allowing for more adaptation. This gives 9 different model parameter $(\beta,p)$ combinations.  

In each TV{--}VAR model $\cM_j$, standard Bayesian forward-filtering analysis applies, and forecasting for daily portfolio rebalancing uses Monte Carlo samples of the predictive distributions for log-prices mapped to the mean vector and covariance matrix on the percent return scale.  Model-specific decisions $\x_{tj}$ use standard Markowitz mean-variance optimization: these minimize predicted portfolio variances among sum-to-one portfolios with daily expected returns constrained to specified target levels $r^*$~\citep[e.g.][sect.~10.4.7]{PradoFerreiraWest2021}.  The initial set of $J=27$ BPDS model/decision pairs is completed by the use of 3 different benchmark targets $r^*\in \{0.05,0.10, 0.15\}$ coupled with each of the 9 TV{--}VAR models.  The actual target return for each model $\cM_j$ on any day $t$ is $r^*_{tj} = \min\{\max(\f_{tj}, 10^{-6}), r^*\}$ where $\f_{tj}$ is the mean vector of predicted returns. This adaptive return target helps down-weight desired returns when the predictions favor small values; this helps to prevent large portfolio positions that would result from using the benchmark target. It also ensures that a portfolio that will lose money is never targeted in expectation. On each day, the resulting optimized portfolio weight vector $\x_{tj}$ is thus a vector of asset weights that has two constraints: they sum to one, i.e., $\x_{tj}'\bone=1$, and they have the specified target  $E_{p_j}(\x_{tj}'\y_t|\cM_j, \D_{t-1}) = r^*_{tj}$ 
for the expected return that day.

\subsection{Selection of Models for BPDS Analysis} 
From the initial set of 27 models, a greedy strategy is used to select a subset of \lq\lq good'' models from analysis over the initial time period 2015-2018. This is to reduce cross-model dependencies in the resulting smaller set and to focus on some of the potentially more useful predictive and decision models. This is done sequentially, reducing from the full model set at each step.  At each step: (i)  consider the set of  remaining models that have historical daily returns whose empirical correlation with those of each of the models already selected are lower than a defined {\em correlation bar} 
 taken as 0.95; (i) among these, 
identify the model with the highest realized 1-day portfolio Sharpe ratio (SR-- computed as realized daily mean return divided by realized daily return standard deviation over the historical data,  multiplied $\sqrt{252}$ to transform to the annual scale); (iii) 
add this model to the selected set until there are no models left with a positive Sharpe ratio or small enough correlation. 

As the models share the same mathematical foundation, there is a high level of cross-model dependence in the initial set so a rather high correlation bar is used to yield several models in the selected set. The bar at 0.95 led to $J=7$ selected models with variability in the defining model order and volatility matrix discount factor parameters.  This suffices for our study; the goal is not necessarily to create an optimal model set that will lead to the highest returns, but rather a well-performing model set for demonstrating aspects and benefits of BPDS in portfolio decisions.

\begin{figure}[htpb!]
    \centering
    \includegraphics[width=\textwidth]{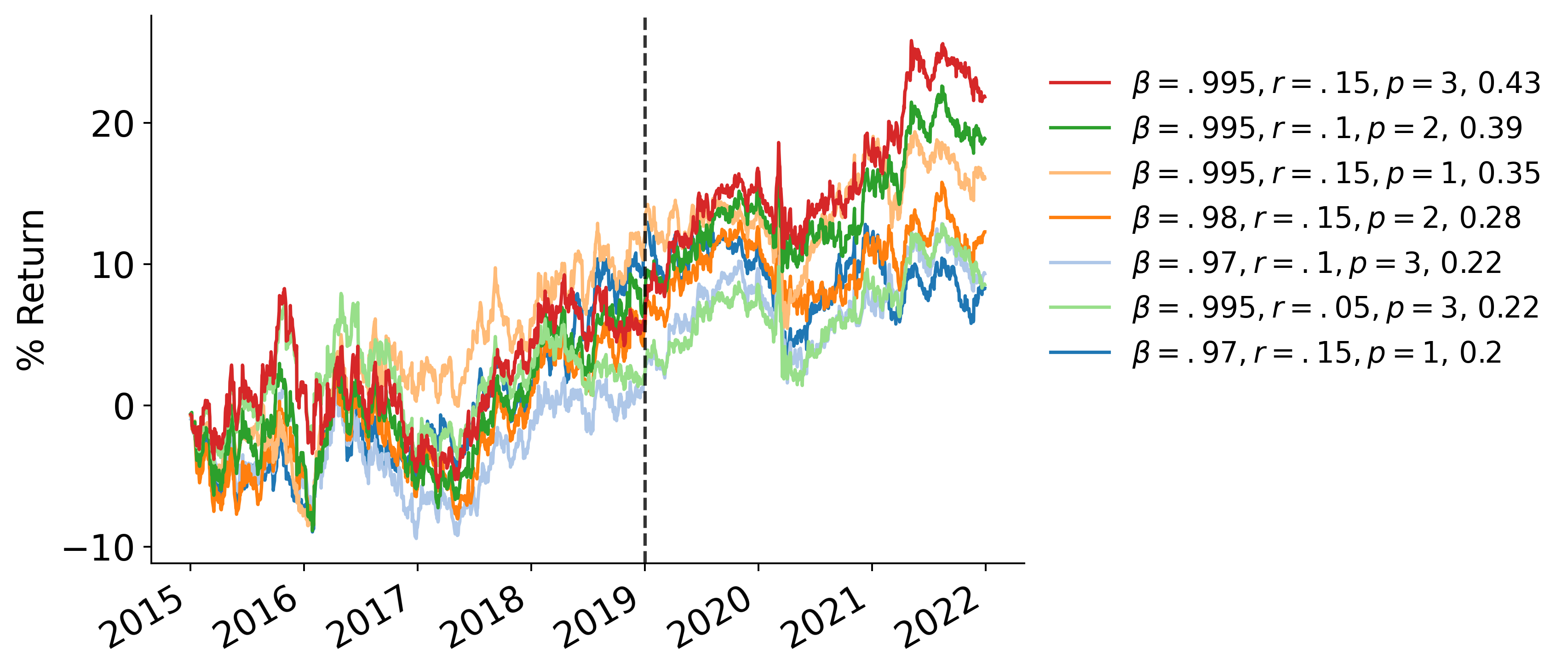}
    \caption{Cumulative portfolio returns over 2015-2018 for the $J=7$ selected models.  
 The empirical SR for each is displayed after the label in the legend indicating the model-specific parameters $(\beta,r=r^*,p)$ .}
    \label{fig:port_mods}
\end{figure}

One relevant detail to note is that the returns for January 15th, 2015 were removed from the analysis. That was the day the Swiss Franc (CHF) was decoupled from the Euro, which resulted in an immediate gain of nearly 30\%  relative to the US dollar. This led to extreme returns across all models, and greatly down-weighted the impact of performance throughout the rest of the test period. To avoid biasing model selection on this single-day event, this day is ignored in the analysis. The selected models, their Sharpe ratios, and maximum return correlations with other selected models over 2015-2018 are shown in Table~\ref{tab:mods}. Cumulative returns and Sharpe ratios from portfolios under each of these models over the entire period are shown in Figure~\ref{fig:port_mods}. 

\begin{table}[htbp!]
\centering
\begin{tabular}{lllllclllll}
 \mc{$\beta$}  &   \mc{$r^*$}   & \mc{$p$}  &  \mc{\em SR} & \mc{$\rho$} 
  & \phantom{.}\qquad\qquad\qquad & 
 \mc{$\beta$} &   \mc{$r^*$}   & \mc{$p$}  &  \mc{\em SR} & \mc{$\rho$} \\ 
\midrule
\ \ 0.995\ \ & \ \ 0.15 \ \ & \ \ 1 \ \  &	\ \ 0.41 \ \ &	\ \ 0.89 \ \ && \ \ 0.98 \ \  & \ \ 0.15 \ \ & \ \ 2 \ \ & \ \ 0.20 &	\ \ 0.95\\
\ \ 0.97  & \ \ 0.15 & \ \ 1 \ \ & \ \ 0.40 \ \ &	\ \ 0.89 \ \  && \ \ 0.995 \ \ & \ \ 0.05\ \  & \ \ 3 \ \ & \ \ 0.11 \ \ &	\ \ 0.95\ \ \\
\ \ 0.995 & \ \ 0.1 \ \ & \ \  2 \ \ & \ \ 0.25 \ \ &	\ \ 0.95 \ \  && \ \ 0.97 \ \  & \ \ 0.1 \ \ & \ \  3 \ \ & \ \ 0.10 \ \ &	\ \ 0.92 \ \ \\
\ \ 0.995 & \ \ 0.15 \ \ & \ \ 3 \ \ &	\ \ 0.22 \ \  &	\ \ 0.95 \ \   && &&&&\\
\end{tabular}
\caption{Annualized values of realized Sharpe ratios (SR) and maximum daily return correlation $\rho$ with other selected models within the set for the $J=7$ models selected based on data over 2015-2018.}
\label{tab:mods}
\end{table}

\section{Initial BPDS Analysis} \label{sec:replicate}

\subsection{Portfolio Setting: Forecast Horizon and Portfolio Targets}
The initial example replicates the set-up of the portfolio example in \cite{TallmanWest2023} but with the new data set and expanded model set. The score function is $\s_{tj}(\y_t, \x_{tj}) = [\x_{tj}'\y_t, -(\x_{tj}'\y_t - \portar_t)^2/2]'$, with further relevant details discussed  in Section~\ref{sec:score_funct} above. The $\portar_t$ in the first dimension is dropped for simplicity and ease of specification for the target score improvement in terms of a percentage.  As $\portar$ is just a constant with respect to $\y_t$ and $\x_{tj}$, this does not change any previous results and has no effect on the resulting analysis. Note that, unlike the settings in~\cite{TallmanWest2023}, $\portar_t$ now depends on $t$,  selected such that  $\portar_t=\sum_{j=0:J} \pi_{tj} r_{tj}^*$, a dynamic target based on the specified targets in the underlying models. This does not affect the mechanics of the BPDS methodology; it simply ensures that BPDS is more adaptive to exploit the fact that the model targets are now changing over time. Several BPDS specifications are evaluated to help understand how BPDS is applied and to explore robustness. 

One main variant explored is the portfolio forecast horizon. With this motivation, results from evaluating both $h=1$ and $h=5$ day ahead portfolios are included. The 1-day portfolios are the portfolios introduced previously, calculated using the predicted next-day returns. The 5-day portfolios are made by fitting each TV--VAR model to daily returns, as in the 1-day case, but then forecasting asset returns 5 days into the future. This gives predictions $p_{tj}(\y_t|\cM_j,\cD_{t-1})$ where now $\y_t$ is redefined as $\y_t = \mathbf{p}_{t+4}/\mathbf{p}_{t-1} - 1$. The moments of this distribution are then used to define 5-day ahead portfolios with the same sum-to-one constraint along with the revised target expected return in $\cM_j$ as
$\min\{\max(\f_{tj}^{(5)}, 10^{-6}), r_{tj}^*\}$ where $\f_{tj}^{(5)}$ is the mean forecast vector 5-days ahead. Though calculated using a 5-day portfolio, these portfolios will still be updated each day; a given portfolio will only be held for a single day before rebalancing. The interest in 5-day ahead analyses is partly contextual in that such portfolios can have better risk profiles and smaller movements in portfolio weights. Also, models might be more easily differentiated, as single-day forecasts tend to be more similar across models while the longer horizon can show more distinctions.  

Further explorations consider ranges of chosen percent target score improvements, with 
BPDS target expected scores set as the element-wise product vector 
$\m_t = \bar{\m}_t*\m^*$, where $\bar\m_t = E_p[\s_{tj}(\y_t, \x_{tj})|\cD_{t-1}]$ and $\m^* =(m_1^*,m_2^*)'$  for all combinations of $m^*_1 \in \{1.05, 1.3\}$ and $m^*_2 \in \{.8, .90, .95, .99\}$. The initial probabilities are set using discounted BMA probabilities~\citep{ZhaoXieWest2016ASMBI}; that is, 
$\pi_{tj} \propto \pi_{t-1,j}^\alpha p_j(\y_{t-1}|\cD_{t-2})$ at each $t$. The discount factor $\alpha = 0.8$ provides discounting of historical model weights between days and hence-- relative to standard BMA which has $\alpha=1$-- analysis avoids the eventual degeneration of model probabilities. 
The BPDS portfolio is calculated using Markowitz optimization, just as in each of the 
models, but now with a target return of $\portar_t + d_t$ with risk tolerance $d_t = \tau_{t1}/\tau_{t2}$. This is the value of $r$ that maximizes $\btau_t'\s_t(r) = \tau_{t1} r - \tau_{t2}/2 (r-\portar_t)^2$, the weighted sum of the score function utilizing BPDS weighting vector $\btau_t$.
We additionally restrict the tilting vector such that $ d_t < \portar_t$ so that the portfolio target can be at most doubled; this is in order to obviate large values earlier seen in Figure~\ref{fig:tau_ratio}, and is enforced in the computations to evaluate $\btau_t$ at each time point. 

\subsection{Some Summary Results}

\begin{figure}[pb!]
    \centering
    \includegraphics[width=.85\textwidth]{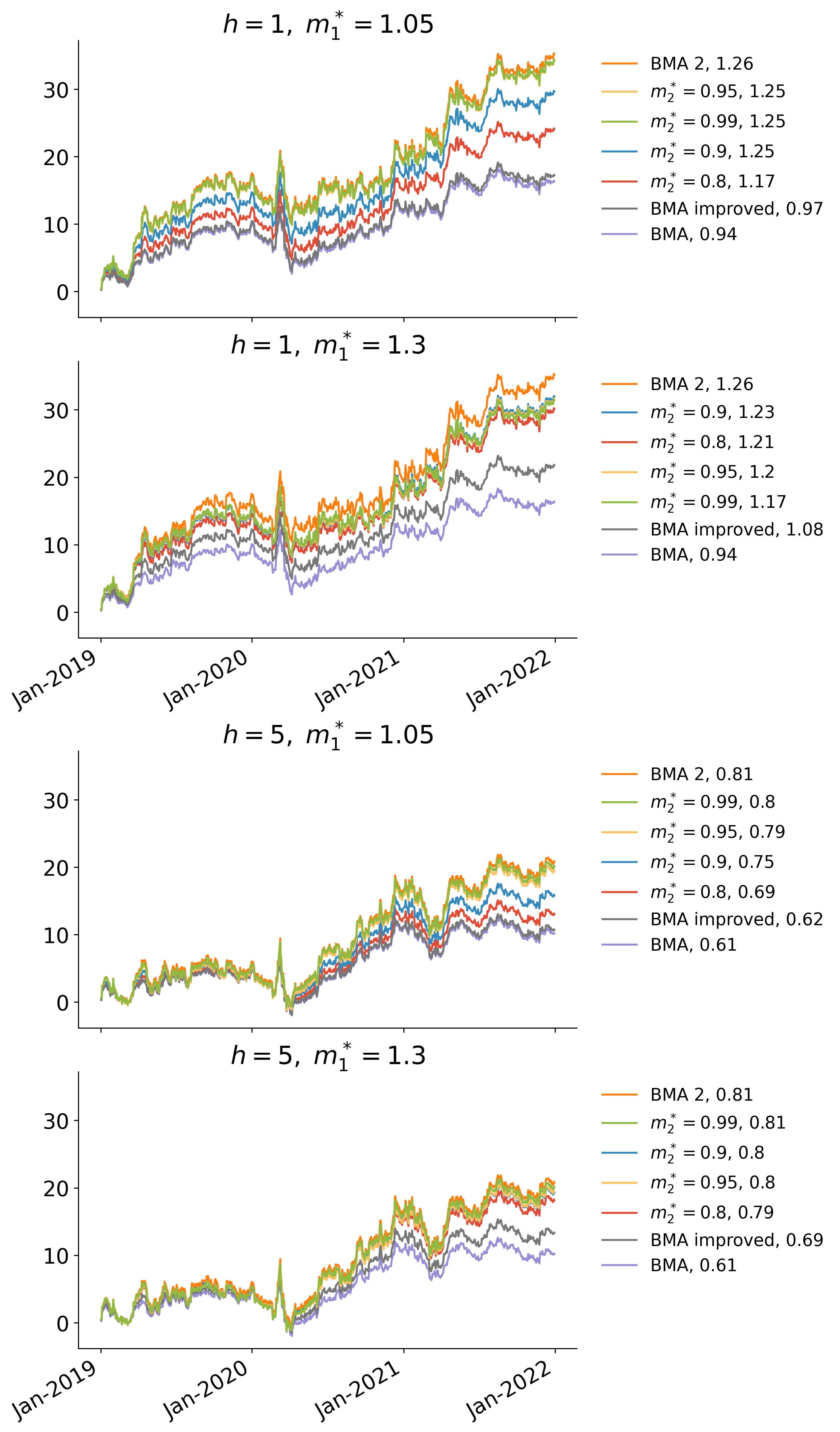}
    \caption{Cumulative returns under BPDS with discounted BMA initial probabilities and return target $\portar_t + \tau_{t1}/\tau_{t2}$ for a variety of score targets and subject to $\tau_{t1}/\tau_{t2} < \portar_t$. This analysis covers the secondary testing period 2019-2021 with portfolio horizon $h=1$ (upper) and $h=5$ (lower). Annualized values of realized Sharpe ratios are displayed after the model labels in the legends.}
    \label{fig:BMA_tau_imp}
\end{figure}

Comparative results involving cumulative returns and Sharpe ratios are seen in Figure~ \ref{fig:BMA_tau_imp}. Additional comparisons come from the standard BMA analysis ($\pi_{t0}=0$, $\btau_t=\bzero$ for all $t$, and target return $\portar_t$). Then, in an effort to advantage BMA through a modified specification of the portfolio setup, 
we added extensions for potential improvements on BMA using target returns $(1+m_1^*)\portar_t$ and $2\portar_t$, labelled as \lq\lq improved BMA" and \lq\lq BMA~2" respectively. These connect to the fact that $m_1^*$ is the desired percent return score improvement in BPDS, and  $2\portar_t$ is the upper bound on $\portar_t + \tau_{t1}/\tau_{t2}$ which defines the target return for BPDS. 
 
BPDS is able to achieve superior cumulative returns and Sharpe ratios compared to  BMA and  \lq\lq improved BMA'', across all settings of $\m_t$. The 5-day portfolios improve over BMA comparisons but generally see relatively lackluster results compared to the 1-day portfolios. This is mainly due to the use of initial BMA-based weights. BMA focuses wholly on 1-step ahead predictive accuracy, so the initial probabilities used here in BPDS generally favour models based on predictive performance at this short-term horizon. This represents a disconnect in BPDS when focused on 5-day portfolios.  A relevant extension would be weighting models using 5-day forecast accuracy/likelihood weightings. Then, there is no obvious pattern of improvement across different values of $m_{t1}^*$ and $m_{t2}^*$, partly due to the restriction placed on $d_t = \tau_{t1}/\tau_{t2}$ that emphasizes control on the risk:reward behavior regardless of the target setting and drives most of the results.

BPDS outcomes are generally close to those of BMA~2, but do not empirically improve upon the latter in this first set of comparisons.  This indicates that most of the performance gains come from the target increase and that the tilting and variation in the target increase are not as helpful. However, note that the methodology for setting the score improvement here is rather simplistic, using a grid of percent improvements. In particular, this does not consider the underlying distribution of the score function which implies, for example, that a 10\% decrease in variance could be a fairly large jump, whereas a 5\% increase in the target return itself is a fairly small improvement.  This observation is key to the methodological extension in the next section that explicitly involves consideration of aspects of the predictive distributions for scores in setting the BPDS targets-- and yields substantial advances in BPDS performance and dominance over the BMA extensions.

\section{Practicable Portfolio BPDS: Structured Score Targets}

\subsection{Score Standardization and Eigenscore Targets} 

Recent use of BPDS in economic applications in~\citet{ChernisTallmanKoopWest2024} introduced target expected score specifications that reflect the importance of relative scales and dependencies among the elements of each $\s_{tj}(\y_t,\x_{tj})$ under the initial distribution $p_t(\y_t,\cM_j|\cD_{t-1})$ at each time $t-1$. We have discussed aspects of this above in Section~\ref{sec:scoredistributions} and revisit it here. A key point is to exploit aspects of the distribution of scores to guide choices of relevant targets that are not wildly inconsistent with the distribution. This aids in avoiding overly aggressive or inconsistent targets such as those highlighted in the naive score targets used in the previous section. 

At each time $t-1,$ write $\bar\m_{tp}$ and $\V_{tp}$ for the mean vector and variance matrix of 
$\s_{tj}(\y_t, \x_{tj}) = [\x_{tj}'\y_t, -(\x_{tj}'\y_t-\portar_t)^2/2]'$ under 
$p_t(\y_t,\cM_j|\cD_{t-1})$.  With  eigendecomposition  
$\V_{tp} = \E_t\D_t^2\E_t'$ where $\D_t = \diag(\delta_{t1},\delta_{t2})$ with positive elements, 
the  {\em eigenscore} $ \D^{-1}\E_t'[\s_{tj}(\y_t, \x_{tj}) - \bar\m_{tp}] $
is a {\em standardized score vector} with zero mean vector and identity variance matrix.  Specified shifts in each dimension of the standardized score vector are now on the same scale and so directly comparable, while the orthogonality implies that there will be limited interaction between them in the ensuing BPDS analysis. Hence it is natural and intuitive to define BPDS analysis based on 
the specification of {\em standardized expected scores} that then map back to the original score scale. That is, with a specified target vector $\m_{t\epsilon}=(m_{t\epsilon1},m_{t\epsilon 2})'$ on the standardized scale  the implied target score on the original scale is
$\m_t = \bar\m_t+ \E_t\D_t\m_{t\epsilon}.$

We know from Section~\ref{sec:scoredistributions} that 
$\V_{tp}$ is typically close to diagonal (and exactly diagonal in some cases) with a much larger variance on the first score element than the second. That is, $\E_t\approx\I$ or exactly $\I,$ and $\delta_{t1}\gg\delta_{t2}$. Hence 
$\m_t \approx$ (or exactly $=$) 
$\bar\m_t+\D_t\m_{t\epsilon} = (m_{t\epsilon 1}\delta_{t1},m_{t\epsilon 2}\delta_{t2})'.$ 
So, with comparable values of the two standardized score targets, the fact that $\delta_{t1}\gg\delta_{t2}$ implies a much greater-- and undesirable-- increase of the target expected return (first element of $\m_t$) than the risk (second element of $\m_t$).    Further insights come from theoretical results in~\citet[][sect.~3.4]{TallmanWest2023} that define approximate expressions for $\btau_t$  when specified target scores are \lq\lq small deviations" from $\m_{tp}.$   In the current setting that theory leads to $\btau_t \approx \D_t^{-1}\m_{t\epsilon} = (m_{t\epsilon 1}/\delta_{t1},m_{t\epsilon 2}/\delta_{t2})'.$  This shows that, for given $\m_{t\epsilon}$ and knowing $\delta_{t1}\gg\delta_{t2}$, there is much greater shrinkage towards zero 
of $\tau_{t1}$ than of $\tau_{t2}.$ This implies that a much larger improvement is needed in the return dimension to achieve a similar level of tilting as in the risk dimension.

An initial use of the standardization concept in this setting explores $\m_{t\epsilon} = \epsilon\bone$ for some small $\epsilon>0$ defining equal increments in each score dimension.  To assess relevant values of $\epsilon,$ we begin with a focus on the target excess return itself.
Set $m_{t1} = \bar{m}_{t1} +\phi|\bar{m}_{t1}|$ for some $\phi$ such that $0<\phi\ll 1$, so that $100\phi$ represents the percent target increase over the initial expected return.  Equating this to $m_{t1} = \bar{m}_{t1}+\epsilon_t \delta_{t1}$   yields the (small) $\epsilon_t = \phi \bar{m}_{t, 1}/\delta_{t1}.$   The implication for the target for the risk element of the score vector is that $\bar{m}_{t2} +\phi \bar{m}_{t1}\delta_{t2}/\delta_{t1}.$  This target score strategy is now used in FX reanalysis. 

\subsection{FX Portfolio Analysis Revisited}

\subsubsection{Example Analysis with Eigenscores}

The analysis of Section~\ref{sec:replicate} is repeated, now setting BPDS score targets with the eigenscore strategy. We compare results under a range of percent improvements $100\phi$ on the expected return. Note that in this setting there is the result that the implied risk tolerance level $d_t = \tau_{t1}/\tau_{t2} \approx \delta_{t2}/\delta_{t1}$. As a result, the BPDS target return improvement will generally be constant across $\phi$ and any improvements for two different values of $\phi$ will be the result of the tilting and not due to different increases in the BPDS target return. 

\begin{figure}[htbp!]
    \centering
    \includegraphics[width=.85\textwidth]{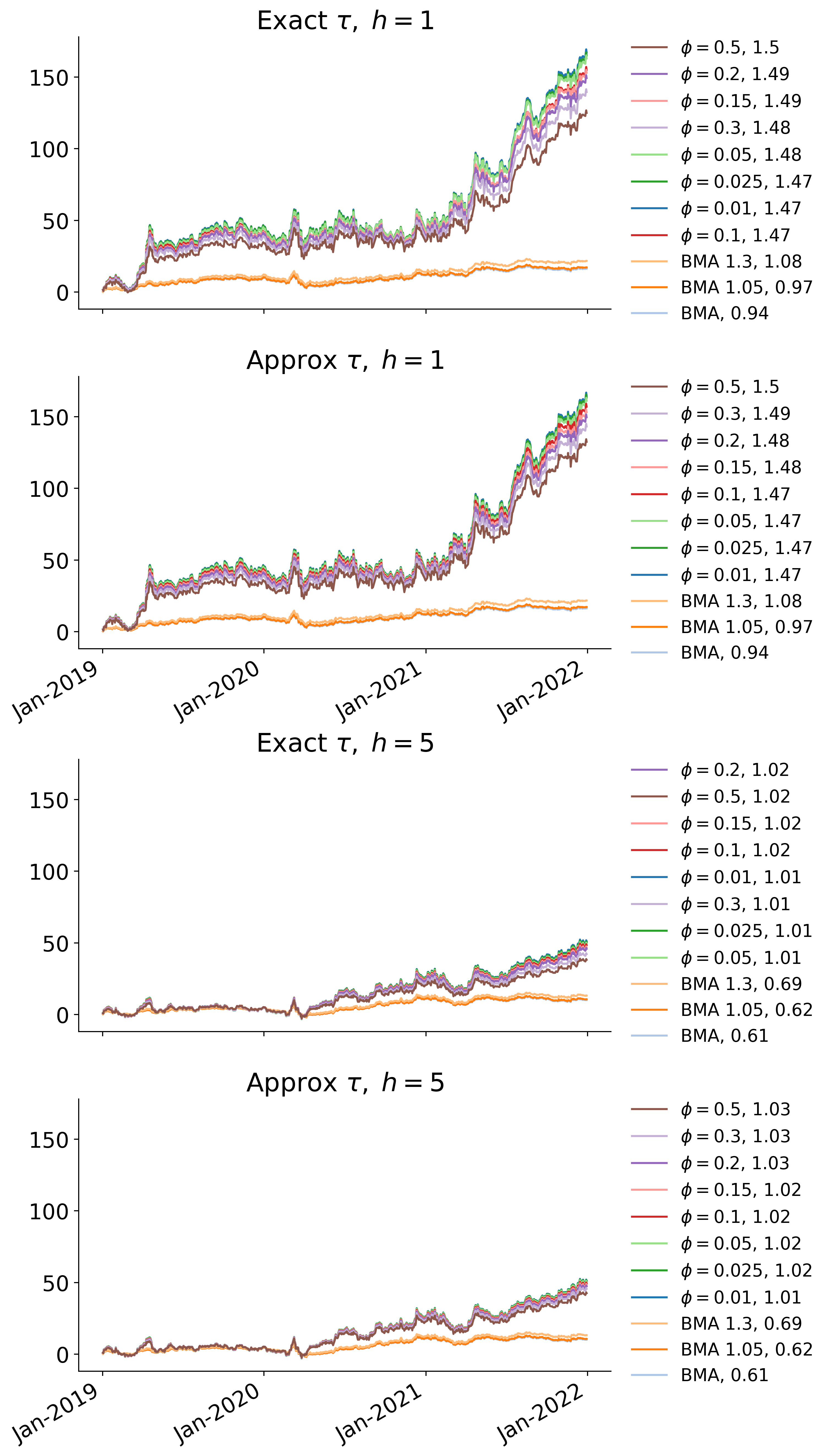}
    \caption{Cumulative returns in BPDS eigenscore analysis for different values of target expected return defined by $\phi.$ Results are from the period 2019-2021, for 1-day (upper) and 5-day (lower) portfolios, and with portfolio return target $\portar_t + \tau_{t1}/\tau_{t2}$. Annualized values of realized Sharpe ratios are displayed after each label in each of the figure legends.}
    \label{fig:BPDS_eig}
\end{figure}

\begin{figure}[htpb!]
    \centering
    \includegraphics[width=\textwidth]{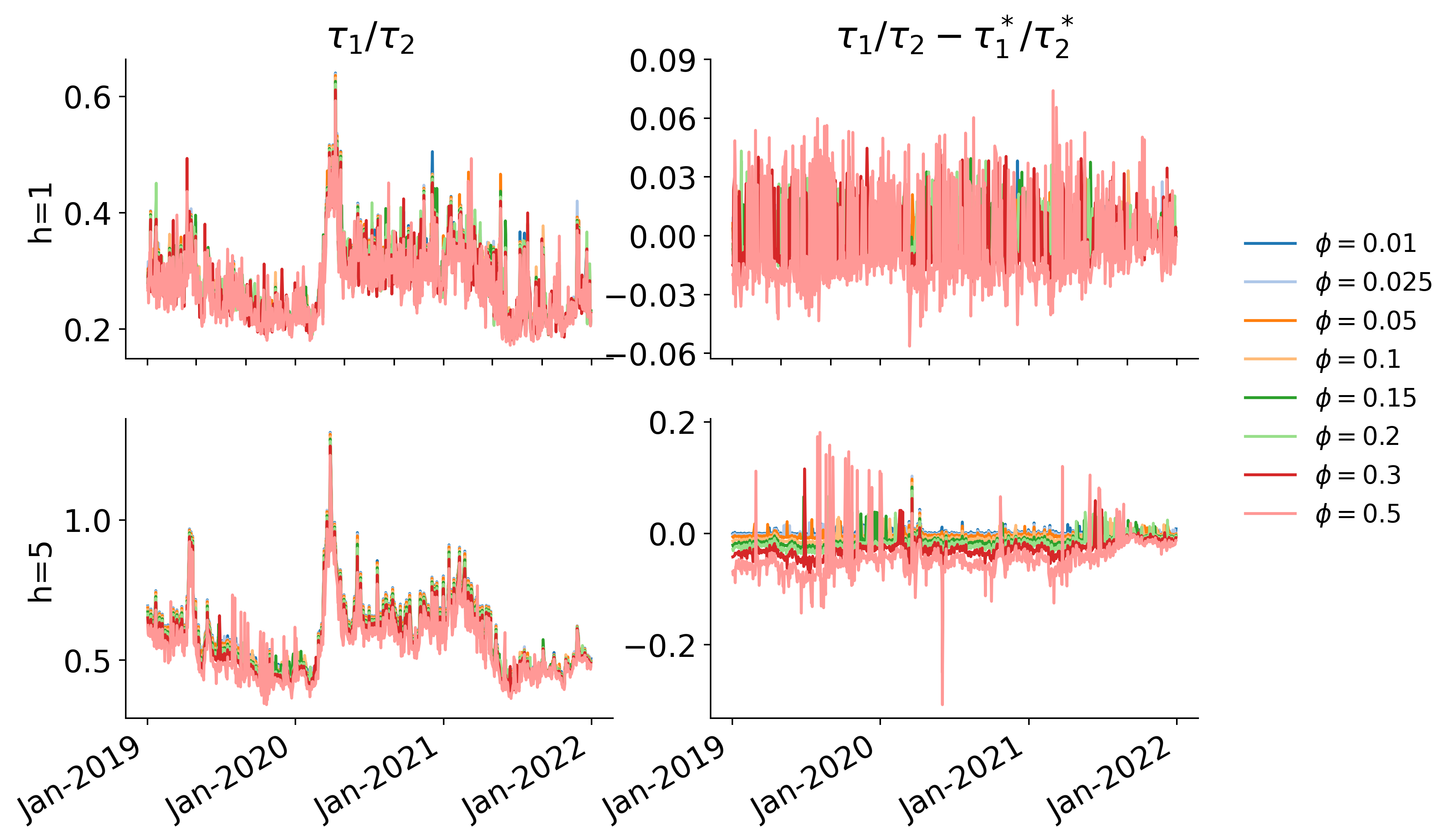}  
    \caption{Return target increase  $\tau_{t1}/\tau_{t2}$ (left) and the difference between the exact $\tau_{t1}/\tau_{t2}$ and the approximation $\tau_{t1}^*/\tau_{t2}^*$ (right) using eigenscore BPDS for a variety of $\phi$.}
    \label{fig:BPDS_eig_target}
\bigskip\bigskip
    \includegraphics[width=\textwidth]{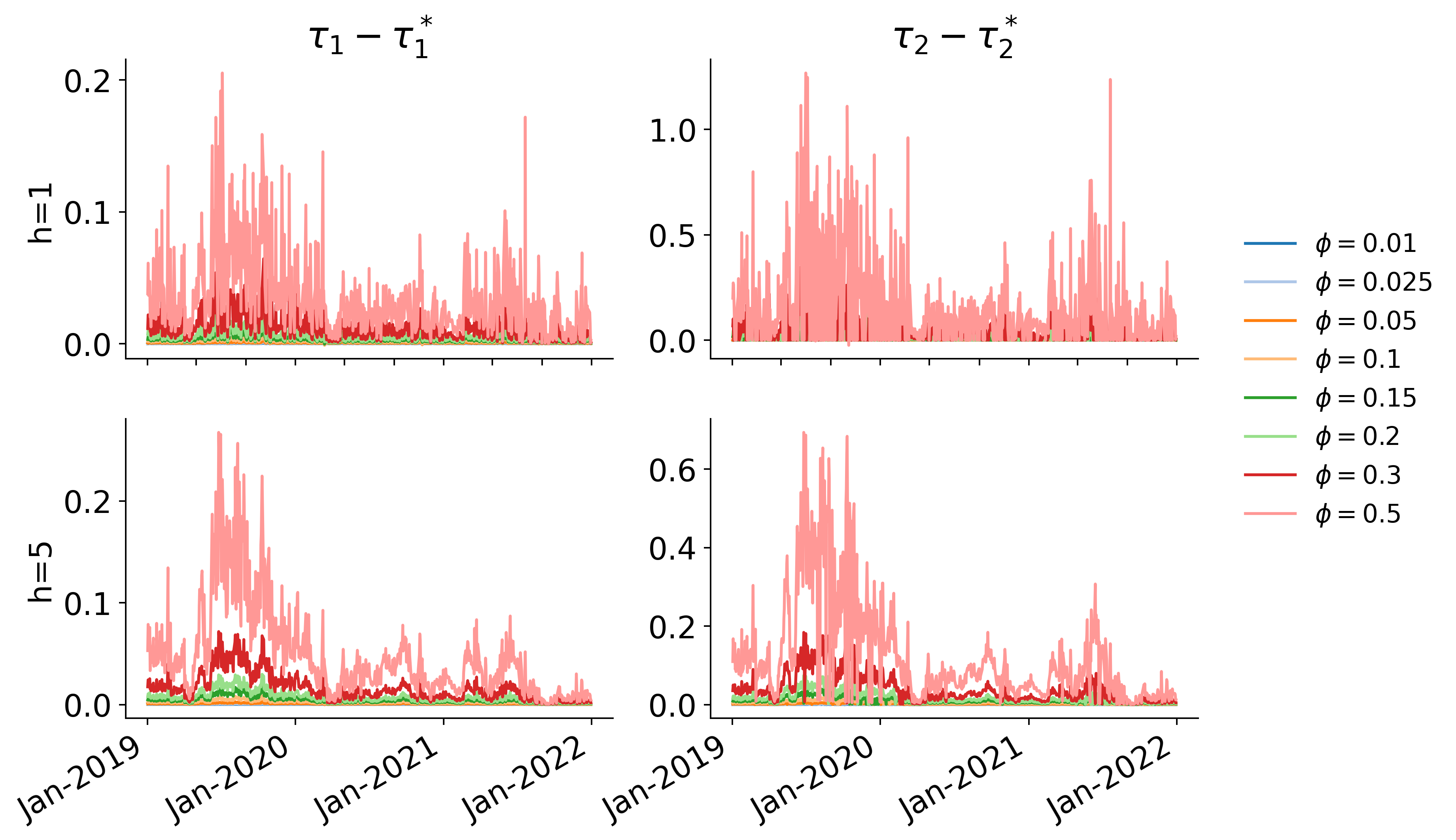}  \caption{Difference between the exact $\btau_t$ and the approximation $\btau_t^*$ using   eigenscore BPDS  for various  $\phi$.}
    \label{fig:BPDS_eig_tau}
\end{figure}

The analysis is repeated for all $\phi \in \{0.01, 0.025, 0.05, 0.1, 0.15, 0.2\}$. 
Cumulative returns and Sharpe ratios are shown in Figure~\ref{fig:BPDS_eig}. Results are shown from analysis using the approximation $\btau_t \approx (m_{t\epsilon 1}/\delta_{t1},m_{t\epsilon 2}/\delta_{t2} )'$ as well as from exact calculations; this aids in empirical investigation of the accuracy of the approximation. Resulting differences in $d_t = \tau_{t1}/\tau_{t2}$ and $\btau_t$ between the approximate and exact analyses are in Figures~\ref{fig:BPDS_eig_target} and \ref{fig:BPDS_eig_tau}.

Portfolio results under all BPDS specifications far exceed those under BMA and the improved BMA approaches, using either the exact or approximate calculations of $\btau_t$. The 1-day portfolios have superior results in terms of both returns and Sharpe ratios, again mainly due to the lack of connection between the initial weights and the BPDS score in the 5-day case.  There are some small differences between the exact and approximate versions for larger values of $\phi$. This is expected as higher $\phi$ correspond to higher values of $\epsilon$, and the theoretical approximation is for \lq\lq small $\epsilon$".  The high increases in returns are partly explained by Figure~\ref{fig:BPDS_eig_target}; the return target is drastically increased, leading to a target of a nearly 0.35\% daily return. This aggression in terms of seeking high returns leads to the extreme returns seen in Figure~\ref{fig:BPDS_eig}.  Since the Sharpe ratios are normalized, this BPDS specification can be realistically compared with BMA despite the difference in return scales, and the results demonstrate realistic and marked improvements using BPDS.

\subsubsection{Repeat Analysis using Modified Target Scores}

The above results are very positive in terms of BPDS improving decision outcomes. However, the large values of risk tolerance $d_t = \tau_{t1}/\tau_{t2}$ generated overly optimistic return targets. Knowing that  $d_t = \delta_{t2}/\delta_{t1}$ in this setting, and that this ratio cannot be changed by altering $\phi$, a modified choice of standardized target score is suggested to address this. This simple modification replaces the standardized target $\epsilon\bone$ with  $\epsilon (1,c)'$ for a chosen constant $c>0$. This factor $c$ represents and induces differences in the degrees of tilting in the return and risk dimension of the score. In particular, a value $c>1$ naturally acts to increase the role of the risk score element and hence can balance an overly optimistic/aggressive return target driving the first element. Some numerical examples below explore this. In these examples,  $c$ is chosen based on the initial test data analysis; its value is determined so that the average over the test time period of generated values of $d_t$ is equal to $d^*$, a chosen average threshold on the risk tolerance measure. 

\begin{figure}[htpb!]
    \centering
    \includegraphics[width=.85\textwidth]{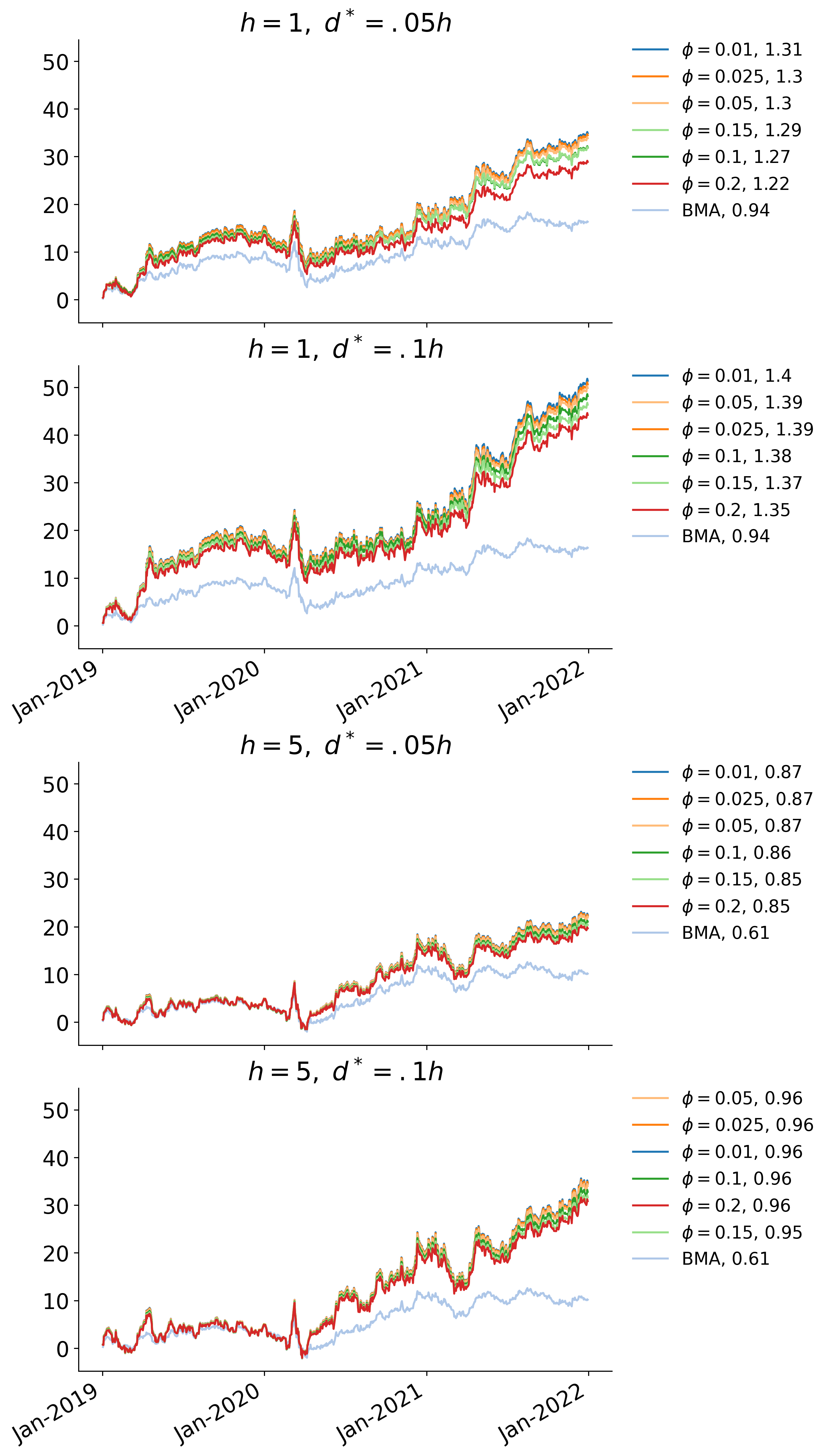}
    \caption{Cumulative returns using standardized target improvement and return/risk upper constraint $d^*$. Results are from the period 2019-2021, for 1-day (upper) and 5-day (lower) portfolios, and with portfolio return target $\portar_t + d_t$ with $d_t=\tau_{t1}/\tau_{t2}$ constrained to be less than $d^*.$   Annualized values of realized Sharpe ratios are displayed after each $\phi$ label in each legend.}
    \label{fig:BPDS_eig_c}
\end{figure}

The examples here are based on choices $d^*=0.05h$ and $d^*=0.1h$, linked to the $h=1$ and $h=5$ day horizons for portfolio targets.  This implies that, on average over the initial test data set, the return target increase will be less than either $0.05$ or $0.1$ for the 1-day portfolio.  This will lead to a more practically reasonable target compared to the aggressive $0.35$ daily target in the analysis of the previous section, an analysis in which the tolerance was unconstrained. This leads to $c=7.4$ and $c=3.7$ respectively when $h=1$, and $c=3.25$ and $c=1.625$ when $h=5$.   Portfolio outcomes shown here are based on analyses using the exact calculation of $\btau_t$; the approximate values are now used only as starting points for the optimization to evaluate $\btau_t$.  Realized cumulative returns and Sharpe ratios are shown in Figure~\ref{fig:BPDS_eig_c}.

Realized cumulative returns are now on a much more practically reasonable scale, while still maintaining a nice risk-adjusted return. All BPDS specifications for the 1-day portfolios are seen to substantially improve over BMA relative to the initial exploration in Section~\ref{sec:replicate}. Note also that the empirical cumulative returns under the 5-day strategy are more comparable with those under the 1-day strategy than in the earlier, unconstrained analysis. That said, there is still room for improvement in the multi-day strategies as they are here still using initial model probabilities based on variants of BMA-based weights. As noted earlier, the latter are wholly based on 1-step forecast accuracy. This is a main question in terms of future development of more appropriate initial model weightings, among which will be to adapt prior goal-focused scoring approaches to refine the specification of the initial $\pi_{tj}$~\citep{LavineLindonWest2021avs,LoaizaMayaJOE2021}.

There is some-- though relatively limited-- variation in empirical returns as $\phi$ is varied, again due almost wholly to differences in resulting tilting. Small values of $\phi$ are empirically preferable in terms of Sharpe ratios, contrasting with the results from the naive analysis of the previous section. This again demonstrates the importance and value of 
\lq\lq small" changes in BPDS target score relative to initial predictions.

Exploration of realized characteristics of portfolio decisions is naturally tied to the time period chosen and displayed. The cumulative return trajectories and Sharpe ratios shown in Figure~\ref{fig:BPDS_eig_c} are for the full 3-year test period.  While not a main point in terms of developing and illustrating the BPDS methodology, it is of contextual interest to look at outcomes over different time periods. Our choice of 2019-2021 inclusive is contextually interesting due to the major differences in economic and market behavior over these years, and some insights are generated by looking at year-specific results. As a snap-shot, Table~\ref{tab:eig_yrs} lists realized Sharpe ratios for each specific year under the 1-day portfolio strategy.  
The main points to note are (i) the year-to-year differences, with 2020 being a stand-out in terms of relatively poor outcomes across models, and (ii) the strong consistency in Sharpe ratios within each year as $\phi$ varies to define different BPDS targets.  

As already emphasized, the empirical returns under BPDS as shown in Figure~\ref{fig:BPDS_eig_c} are substantially superior under any of the BPDS variants relative to BMA.  That the Sharpe ratios per year under BMA are relatively competitive is due to the fact that  BPDS spreads the model probabilities more than BMA each time point, and always has some probability on the over-diffuse baseline to address the model-set incompleteness question.  Hence BPDS will almost surely define heavier-tailed forecast distributions for $\y_t$ that lead to higher uncertainties on portfolios than under BMA. The latter  will typically underestimate uncertainties as it concentrates quickly around one of the initial $J$ models. This helps in interpreting the realized Sharpe ratios that represent just one aspect of \lq\lq performance", and that is balanced by the realized return outcomes under BPDS as shown in Figure~\ref{fig:BPDS_eig_c} to emphasize this point in this case study. Here, under BPDS the appropriately increased levels of uncertainty are balanced by substantially increased realized returns.  Additional metrics highlighting differences between the strategies-- including purely predictive comparisons in the full space of returns defined by models for $\y_t$-- will be of interest in further comparisons.

\begin{table}[pt!]
    \centering
\begin{tabular}{@{}l|ccc|ccc@{}}
               & \multicolumn{3}{c|}{\textbf{$d^*=0.05$}}                   & \multicolumn{3}{c}{\textbf{$d^*=0.10$}}                       \\ 
           & 2019           & 2020            & 2021           & 2019            & 2020           & 2021           \\\midrule
$\phi= 0.01$ \quad  & \ \  2.21   \ \         & \ \  0.21  \ \           & \ \  1.98  \ \         & \ \  2.14  \ \         & \ \  0.16  \ \         & \ \  2.34    \ \        \\
$\phi=  0.025$ \quad  & 2.21 & 0.21          & 1.97          & 2.14          & 0.15          & 2.33          \\
$\phi=  0.05$  \quad & 2.19          & 0.22          & 1.97          & 2.14           & 0.16          & 2.33          \\
$\phi=  0.1$   \quad & 2.21          & 0.21           & 1.89          & 2.14           & 0.18          & 2.29          \\
$\phi=  0.15$ \quad  & 2.20          & 0.26 & 1.90          & 2.12          & 0.16          & 2.28          \\
$\phi=  0.2$   \quad & 2.21          & 0.13            & 1.84          & 2.16           & 0.14          & 2.24          \\ 
BMA          \quad  &  2.17   & 0.23         & 0.88   &  2.17 & 0.23 & 0.88        \\
\end{tabular}
    \caption{Annual values of realized Sharpe ratios using standardized target improvement and return/risk constraint $d^*$ for $h=1$. }
    \label{tab:eig_yrs}
\end{table}

\section{Summary Comments}
 BPDS is a foundational framework that enables the integration of expected and historical decision outcomes in predictive model uncertainty settings. As exemplified here, BPDS can potentially improve realized decision outcomes and thus serve as an important tool for portfolio managers for model combination, calibration, and evaluation. The case study developed in this paper demonstrates the potential for BPDS to improve the trade-off between risk and reward in portfolio analysis. It also provides additional methodology for understanding relevant score functions and setting target scores in the portfolio setting. The results further confirm and expand on the positive results originally reported in \cite{TallmanWest2023}.  The focused development of customized BPDS target scores, and the practical results in terms of desirable portfolio characteristics,    demonstrate the importance of incorporating information about the underlying score distribution, and of the contextual interpretation of the elements of the score vector. The examples show the potential for BPDS to improve on current Bayesian model averaging methods in portfolio analysis, with the prospects for more adaptive portfolios that have better resulting portfolio outcomes.

 There are various potential future directions for applying BPDS in finance. One is in expanding the set of score functions to include other relevant metrics, such as portfolio churn or skewness/kurtosis of the return distribution, and of course other features of portfolio \lq\lq paths" over multiple forecast horizons~\citep[e.g.][]{IrieWest2018portfoliosBA,TallmanPhD2024}.   
Further investigation of the questions related to setting BPDS target scores is also relevant, especially towards improved understanding of how small changes in the target score lead to changes in the resulting decisions. It is also of interest to consider utilizing initial weights that are customized and potentially more relevant than those based on simple, following prior literature on formal Bayesian model weighting based on historical outcomes with respect to specific predictive and decision goals~\citep[e.g.][and references therein]{Karlsson2007, LavineLindonWest2021avs, BernaciakGriffin2024,ChernisTallmanKoopWest2024}.

\subsubsection*{Acknowledgements}
The research reported here was performed while Emily Tallman was a PhD student in Statistical Science at Duke University. Tallman's research was partially supported by the U.S. National Science Foundation through NSF Graduate Research Fellowship Program grant DGE 2139754. Any opinions, findings, and conclusions or recommendations expressed in this material are those of the authors and do not necessarily reflect the views of the 
National Science Foundation.

\small
\bibliographystyle{spbasic}  
\bibliography{TallmanWest}
\end{document}